\documentclass[a4paper,11pt]{article}

 \usepackage{amsmath,amsfonts,amsthm,amssymb}
 \usepackage{latexsym,epsf}

\newcommand{\eq}{\begin{equation}}
\newcommand{\en}{\end{equation}}
\newcommand{\eqa}{\begin{eqnarray}}
\newcommand{\ena}{\end{eqnarray}}

\newcommand{\B}{\mathcal{B}}
\newcommand{\PP}{\mathcal{P}}
\newcommand{\Ss}{\mathcal{S}}
\newcommand{\N}{\mathbb{N}}
\newcommand{\C}{\mathbb{C}}
\newcommand{\bE}{\mathbf{E}}
\newcommand{\Z}{\mathbb{Z}}

\newcommand{\Aut}{\textrm{Aut}}

\newcommand{\Out}{\textrm{Out}}
\newcommand{\Inn}{\textrm{Inn}}
\newcommand{\1}{1\!\! 1}
\numberwithin{equation}{section}
\newtheorem{thm}[equation]{Theorem}

\newtheorem{prop}[equation]{Proposition}
\theoremstyle{definition}
\newtheorem{remark}[equation]{Remark}

\newtheorem{definition}[equation]{Definition}

\begin{document}

\setlength{\unitlength}{1mm} \thispagestyle{empty}

\vspace*{0.1cm}

 \begin{center}
 {\bf  Extraspecial Two-Groups, Generalized Yang-Baxter
Equations and Braiding Quantum Gates
  \\[2mm]}

\vspace{.5cm}
Eric C. Rowell ${}^{a,}$\footnote{rowell@math.tamu.edu},
Yong Zhang ${}^{b,}$\footnote{zhangyo2008@gmail.com. The present address:
Center for High Energy Physics, Peking University, Beijing 10087, China},
Yong-Shi Wu ${}^{b,}$\footnote{wu@physics.utah.edu},
and Mo-Lin Ge ${}^{c,}$\footnote{geml@nankai.edu.cn} \\[.5cm]

{\small  ${}^a$ Department of Mathematics,
  Texas A\&M University, College Station, TX 77843

${}^b$ Department of Physics and Astronomy,  University of Utah,
   Salt Lake City, UT 84112

 ${}^c$ Chern Institute of Mathematics,
  Nankai University, Tianjin 300071, China
\\[0.1cm] }

\end{center}

\vspace{0.2cm}

\begin{center}
\parbox{12cm}{
\centerline{\small  \bf Abstract} \small \noindent

In this paper we describe connections among extraspecial 2-groups,
unitary representations of the braid group and multi-qubit braiding
quantum gates. We first construct new representations of extraspecial
2-groups. Extending the latter by the symmetric group, we construct
new unitary braid representations, which are solutions to generalized
Yang-Baxter equations and use them to realize new braiding quantum
gates. These gates generate the GHZ (Greenberger-Horne-Zeilinger)
states, for an arbitrary (particularly an \emph{odd}) number of qubits,
from the product basis. We also discuss the Yang-Baxterization of the
new braid group representations, which describes unitary evolution of
the GHZ states. Our study suggests that through their connection with
braiding gates, extraspecial 2-groups and the GHZ states may play an
important role in quantum error correction and topological quantum
computing.}


\end{center}

\vspace*{10mm}
\begin{tabbing}

\end{tabbing}

\newpage

\section{Introduction}

Recently, the study of braiding quantum gates has attracted much
attention \cite{dye,kauffman1,yong1,yong2,frw,zkw,yong3,yong4}.
The Bell matrix generating all the Bell states from the product
basis (or the computational basis) was recognized \cite{dye} to
be a four-by-four solution to the braided Yang-Baxter equation
(YBE) \cite{yang,baxter}; later it was shown to provide a
universal (two-qubit) quantum gate \cite{kauffman1}.  Higher
dimensional solutions to the YBE, called the generalized Bell
matrices in \cite{yong4}, were used to introduce the braiding
quantum gates that generate the Greenberger-Horne-Zeilinger
(GHZ) states \cite{ghz1} for an \emph{even} number of qubits
from the product basis. In addition, Yang--Baxterization
\cite{jones1,gxw} of the generalized Bell matrix has been used
\cite{yong4} to derive a Hamiltonian that describes unitary
evolution of the GHZ states of an even number of qubits.

The GHZ states are known to be the simplest multipartite maximally
entanglement sources and have been widely exploited in quantum
information theory (see, e.g. \cite{ghz2,ghz3,nielson}). In
particular, the GHZ states may act as ancillas in fault-tolerant
operations \cite{aharonov}. In this paper we study some algebraic
and topological aspects of the braiding quantum gates that
generate the GHZ states, for an \emph{arbitrary} (particularly
an \emph{odd}) number of qubits. The motivation is to help
establish connections between the GHZ states and topological
quantum information processing and topological quantum computation,
in which the braiding quantum gates are known to play a pivotal
role \cite{kauffman1,kitaev,fklw,kauffman,kl}. Essentially our
present paper presents generalization of a recent paper by
Franko, Rowell and Wang \cite{frw}. In that paper the images of the
unitary braid group representations generated by the $4\times 4$
Bell matrix, corresponding to a two-qubit braiding quantum gate
\cite{kauffman1}, have been identified through extension of
representations of the extraspecial 2-groups. The decomposition
of reducible braid group representations into their irreducible
constituents have also been determined, and it has been shown to
be closely related to the well-known Jones representations
\cite{jones2,jones3} at $4$th roots of unity, associated with
the $SU(2)_2$ model for topological matter states (see, e.g.
\cite{DSetal}). In the present paper we generalize these results
to the construction -- again through extraspecial 2-groups -- of
higher dimensional unitary braid group representations that give
rise to the braiding quantum gates acting on an arbitrary
(particularly an \emph{odd}) number of qubits and generating
multi-qubit GHZ states from the product basis. We also carry out
the Yang-Baxterization of the new braid group representations,
which can be used to describe unitary evolution of the GHZ
states of any number of qubits. Possible connections with
quantum error correction and topological quantum computation
will be discussed in the concluding section.


This paper is organized as follows. In Section 2 we record our
notational conventions and introduce the relevant braid groups
and a generalized form of the Yang-Baxter Equation. Section 3
contains the main mathematical results describing how one
obtains unitary representations of braid groups from those of
extraspecial 2-groups and how the latter may be constructed from
almost-complex structures. In Section 4 we explain how to obtain
GHZ states from the product (computational) basis using the braid
representations found in Section 3, as well as their unitary
evolution under the Hamiltonians derived via Yang-Baxterization
\cite{jones1,gxw}. Section 5 is devoted to concluding remarks on
relevance of extraspecial 2-groups and the GHZ states to quantum
error correction and topological quantum computation, while the
Appendix describes for interested readers a generalized version
of the quantum Yang-Baxter equation developed in this paper.

 \section{Preliminaries}\label{defs}

We use  $1\!\! 1$ to denote the identity operator and
 $1\!\! 1_m$ the $m\times m$ identity matrix. The superscripts
$\ast$ and $\dag$, respectively, denote the complex conjugation and
Hermitian operation of a matrix (or a complex number). The symbol
 $\delta_{ij}$ is the Kronecker function of two variables $i,j$,
 which is $1$ if $i=j$ and $0$ otherwise. The
 function $\epsilon(k)$ is defined by \eq  \label{step}
 \epsilon(k)\epsilon(k)=1,\quad
\epsilon(k)\epsilon(\bar k)=-1, \en which has solutions $
\epsilon(k)=\pm 1$, $\epsilon(\bar k)=\mp 1$, and either ${\bar
k}=-k$ or $\bar k=2 n -k + 1$ depending on the convention used
for the range of $k$ or $\bar{k}$. The tensor product $A\otimes B$
of the matrices $A=(A_{ij})$ and $B=(B_{kl})$ is defined by the
convention $(A\otimes B)_{ik,jl}=A_{ij}B_{kl}$.

 The symbol $A^{J_1J_2}$ denotes a matrix having the following matrix
 entries, \eq
 (A^{J_1J_2})_{\mu a,\nu b},\,\,
 \mu,\nu=J_1,J_1-1,\cdots,-J_1,\,\,
 a, b=J_2,J_2-1, \cdots, -J_2,
\en where  $J_1, J_2$ are integers or half-integers. The matrix
$A^{J_1J_2}$ has the following operator presentation, also denoted as
$A^{J_1J_2}$, in terms of Dirac kets and bras:
 \eq
 A^{J_1J_2}=\sum_{\mu=-J_1}^{J_1}\sum_{\nu=-J_1}^{J_1} \sum_{a=-J_2}^{J_2}
 \sum_{b=-J_2}^{J_2} (A^{J_1J_2})_{\mu a, \nu b} |\mu a\rangle \langle \nu b|,
 \en
where the kets $|\{\mu\rangle\}$ or $\{|a\rangle\}$ form an orthonormal basis,
respectively, in $2J_1+1$ and $2J_2+1$ dimensional Hilbert space:
\eq
 \sum_{\mu=-J_1}^{J_1} |\mu\rangle \langle \mu| =1\!\!
 1_{(2J_1+1)}, \quad
  \sum_{a=-J_2}^{J_2} |a\rangle \langle a| =1\!\! 1_{(2 J_2+1)},
\en where  $\langle \mu | \nu\rangle=\delta_{\mu\nu}$ and  $\langle
a | b \rangle=\delta_{ab}$.

The Pauli matrices
 $\sigma_x, \sigma_y, \sigma_z$ have the conventional forms
 \eq
 \sigma_x= \left(\begin{array}{cc} 0 & 1\\ 1 &0\end{array}\right),
\qquad
\sqrt{-1}\sigma_y=\left(\begin{array}{cc}0 & 1\\ -1 &0
\end{array}\right),\qquad
\sigma_z=\left(\begin{array}{cc} 1 & 0\\ 0 &-1 \end{array}\right).
 \en

Artin's \emph{braid group} $\B_n$ on $n$ strands has the well-known
presentation in terms of generators $b_1,\ldots,b_{n-1}$
satisfying the far-commutation relation
\eq \label{br1}
b_{i} b_{j} = b_{j} b_{i},\qquad |i - j | \ge 2
\en
and the braid relation
 \eq \label{br2}
   b_{i}  b_{i+1} b_{i} = b_{i+1} b_{i} b_{i+1},
   \qquad 1 \leq i \leq n-2.
   \en

$\B_n$ has a finite-index normal subgroup $\PP_n$ generated by
the conjugacy class of $(b_1)^2$. $\PP_n$ is called the
\emph{pure braid group} and can be understood as the kernel of the
surjective homomorphism $\B_n\to\Ss_n$ onto the symmetric group on
$n$ letters sending the generators $b_i$ of $\B_n$ to transpositions
$(i\,i+1)$. This induces an isomorphism ${\mathcal S}_n$ $\cong$
${\mathcal B}_n/{\mathcal P}_n$.  The term ``pure braids" applied
to elements of $\PP_n$ is due to the fact that in the geometric
formulation of $\B_n$ as braiding operators on $n$ strands, the
elements of $\PP_n$ are exactly those that do not permute the strands.

Relation (\ref{br2}) leads to the (braided version of the)
YBE, i.e., \eq \label{ybe}
  (\check{R}\otimes 1\!\! 1_d) (1\!\! 1_d \otimes \check{R})
  (\check{R}\otimes 1\!\! 1_d)=(1\!\! 1_d \otimes {\check R})
 (\check{R}\otimes 1\!\! 1_d)(1\!\! 1_d \otimes \check{R}), \en
with an invertible $d^2\otimes d^2$ matrix $\check{R}$: $V\otimes V
\to V\otimes V$ where $V= \C^d$. The relation (\ref{ybe}) gives rise
to a sequence of representations $(\pi_n, (\C^d)^{\otimes n})$ of
${\mathcal B}_n$:
\eq \pi_n(b_i)=1\!\! 1_d^{\otimes i-1} \otimes \check{R}\otimes
1\!\! 1_d^{\otimes n-i-1} \en since clearly $\pi_n(b_i)$ and
$\pi_n(b_j)$ commute for $|i-j|\geq 2$. This type of braid group
representations was used in \cite{yong4} to construct the GHZ
states for an even number of qubits.

To construct the GHZ for an {\it odd} number of qubits, we will
explore solutions to a {\it generalized version} of the YBE:
\begin{definition}\label{GYBE}
 Fix $p$ with $2\leq p \in \N $ and let $l=p^k$. An
invertible $p^N\times p^N$ matrix $\check{R}$ is a solution to the
\emph{generalized Yang-Baxter Equation} if
\eq
\label{gybe}    (\check{R}\otimes 1\!\! 1_l) (1\!\! 1_l \otimes
\check{R}) (\check{R}\otimes 1\!\! 1_l)  =
 (1\!\! 1_l \otimes {\check R})(\check{R}\otimes 1\!\! 1_l)
 (1\!\! 1_l \otimes \check{R}), \en
  as operators on $(\C^p)^{\otimes (k+N)}$.
 \end{definition}
  When $k=1,N=2$, the
generalized YBE (\ref{gybe}) is the same as the conventional YBE
(\ref{ybe}). If $k\geq N/2$, the assignment
\eq
\pi_n(b_i)=1\!\! 1_l^{\otimes i-1} \otimes \check{R}\otimes 1\!\!
1_l^{\otimes n-i-1} \en defines a sequence of representations
$(\pi_n, (\C^p)^{\otimes (N+k(n-2))})$ of the braid group
${\mathcal B}_n$.  Here relation (\ref{br2}) is clearly satisfied
by the $\pi_n(b_i)$ but relation (\ref{br1}) necessitates the
restriction $k\geq N/2$ as we will see later. In the following we are
going to consider only the case with $p=2$, corresponding to qubits.

Finally we will use the term \emph{almost-complex structure} to
mean a matrix $M$ such that $M^2=-\1$.

\section{Extraspecial 2-groups
and unitary braid group representations}

In \cite{frw}, extraspecial 2-groups are seen to play a key role in
studying the images of the braid group $\B_n$ under the
representation associated with the $4\times 4$ Bell matrix. Inspired
by this work, we present an approach to the GHZ states (higher
dimensional generalizations of the Bell states) starting from
extraspecial 2-groups and their anti-Hermitian representations in
terms of almost-complex structures.

 \subsection{Extraspecial 2-groups}\label{es2groups}

We give a brief sketch of extraspecial 2-groups following \cite{frw}.
The group $\bE_m$ is the abstract group generated by $e_1,\ldots,e_{m}$
with relations: \eq \label{extra} e_i^2=-1\!\! 1,\,\,
e_ie_j=e_je_i,\,\, |i-j|\geq 2,\quad e_{i+1}e_i=-e_ie_{i+1}, \,\,
1\leq i,j\leq m-1, \en Here $-1\!\! 1$ is an order 2 central
element, and we denote $-\1 a$ by $-a$.  Any element in $\bE_m$ can be
expressed in a unique normal form:
$\pm e_1^{\alpha_1}\cdots e_{m}^{\alpha_{m}}$ where $\alpha_i\in\Z_2$,
and $\bE_m/\{\pm 1\!\! 1\}\cong (\Z_2)^{m}$.  It is clear from this
normal form that $\bE_m$ has order $2^{m+1}$.

A group $G$ of order $2^{m+1}$ with $m$ even is an \textbf{extraspecial
2-group} if both the center $Z(G)$
($=\{g\in G: ga=ag\; \mathrm{for all}\; a\in G\}$)
and the commutator subgroup $G^\prime$ (generated by elements of the form
$aba^{-1}b^{-1}$) are isomorphic to $\Z_2$ and $G/Z(G)\cong (\Z_2)^m$
(see \cite{g}). The commutator subgroup of $\bE_m$ is $\{\pm 1\!\! 1\}$
due to its (anti-)commutation relations, and hence it is immediate
that $\bE_{2k}$ is an extraspecial 2-group.  When $m$ is odd the center
of $\bE_m$ has order 4.  However, since
$\bE_{m-1}\subset\bE_m\subset\bE_{m+1}$ we obtain an extraspecial 2-group
from $\bE_m$ by adding or removing a generator, so we will call $\bE_m$
with $m$ odd a \emph{nearly} extraspecial 2-group.

The center of $\bE_m$ is
 \eq \label{center} Z(\bE_m)=\left\{\begin{array}{cc} \{\pm 1\!\! 1\} &
\textrm{$m$ even}\\
\{\pm 1\!\! 1, \pm e_1e_3\cdots e_m\} & \textrm{$m$ odd}
 \end{array}\right.
 \en
For $m=2 k-1$ odd, the form of the center of $\bE_{2k-1}$ depends
on the parity of $k$:
  \eq Z(\bE_{2k-1})
   \cong  \left\{ \begin{array}{cc} \Z_2\times\Z_2 & \textrm{$k$ even}\\
\Z_4 & \textrm{$k$ odd} \end{array}\right. \en

\subsection{Irreducible representations of $\bE_m$}\label{esreps}

Irreducible representations of $\bE_{m}$ are described in detail
in \cite{frw}.  We summarize them here to establish notation, for
explicit realizations, see \cite{frw}.  For $m=2k$, there are a
unique irreducible $2^k$-dimensional representation which we
denote by $(\rho_1,V_1)$, and $2^{2k}$ inequivalent 1-dimensional
representations $(\rho_j, V_j)$, $j=2,\cdots,{2^{2k}+1}$ of the
form
  \eq
\rho_j(\pm 1\!\! 1)=1, \quad \rho_j(\pm e_i)=\pm 1,\,\,
 i=1,\cdots 2k .
  \en
Note that in the right hand side of the second equation, for a
fixed $i$ there are two choices of the sign; altogether there are
$2^{2k}$ choices, corresponding to $2^{2k}$ inequivalent
representations labelled by $j=2,\cdots,{2^{2k}+1}$.

For $m=2k-1$, $\bE_m$ has $2^{2k-1}$ inequivalent 1-dimensional
representations denoted by $(\lambda_j,W_j)$,
$j=3,\cdots,2^{2k-1}+2$ of the form \eq \lambda_j(\pm 1\!\! 1)=1,
\quad \lambda_j(\pm e_i)=\pm 1,\,\, i=1,\cdots 2k-1. \en A note
similar to that below eq. (3.4) applies here. Moreover,
$\bE_{2k-1}$ has two inequivalent irreducible
$2^{k-1}$-dimensional representations $(\lambda_1,W_1)$ and
$(\lambda_2,W_2)$.

The following key proposition shows that
certain representations of $\bE_{n-1}$ induce representations of $\B_n$:
\begin{prop}\label{es2br}
Let $\{T_1,\ldots,T_{n-1}\}$ be a set of $k\times k$ matrices such that:
\begin{enumerate}
\item[(E1)] $T_i^2=-\1$,
\item[(E2)] $T_iT_j=T_jT_i$ if $|i-j|>1$,
\item[(E3)] $T_iT_{i+1}=-T_{i+1}T_i$ for all $1\leq i\leq (n-2)$.
\end{enumerate}
Then \begin{enumerate}
\item[(a)] The assignment $\phi_{n-1}(e_i)=T_i$ defines a representation
of $\bE_{n-1}$. Moreover, $\phi_{n-1}$ contains no 1-dimensional
subrepresentations.
\item[(b)] Let
$\check{R}_i=\frac{1}{\sqrt{2}}(\1+T_i)$, $1\leq i\leq n-1$.
Then $b_i\rightarrow \check{R}_i$ defines a representation of $\B_n$.
 \item[(c)] If in addition
the $T_i$ are anti-Hermitian (i.e. $T_i=-T_i^\dag$), the $\B_n$
representation is unitary.
\item[(d)] Moreover, if $n$ is odd, $\phi_{n-1}$ is a direct sum
of $\frac{\dim(\phi_{n-1})}{2^{(n-1)/2}}$ copies of $\rho_1$.
\end{enumerate}
\end{prop}
\begin{proof}
Since the $T_i$ satisfy (\ref{extra}) it is clear that
$\phi_{n-1}$ does define a representation of $\bE_{n-1}$. If the
central element $-\1\in\bE_{n-1}$ has image $-\1$ then this
holds for any subrepresentation of $\phi_{n-1}$ as well. By the
explicit construction of 1-dimensional representations of
$\bE_{n-1}$ in \cite{frw}, $\phi_{n-1}$ has no 1-dimensional
representations since $-\1$ acts by $1$ for these representations.
This proves (a). Claim (d) follows immediately from (a), since for
$n-1$ even there is only one irreducible representation of
dimension more that 1, and its dimension is $2^{(n-1)/2}$.

The matrices $\check{R}_i$ obviously satisfy relation (\ref{br1})
since the $T_i$ satisfy $(E2)$. Moreover, we have:
\begin{eqnarray*}
\sqrt{2}\check{R}_i\check{R}_{i+1}\check{R}_i&=&T_iT_{i+1}T_i+2T_i
  +T_{i+1}+T_iT_{i+1}+T_{i+1}T_i+T_i^2+\1\\
&=&2T_{i+1}+2T_i+T_iT_{i+1}+T_{i+1}T_i=2T_{i+1}+2T_i,
\end{eqnarray*}
which is symmetric under $i\leftrightarrow i+1$ so that the
$\check{R}_i$ satisfy the braid relation (\ref{br2}), proving (b).

Observing that $\check{R}_i^{-1}=\frac{1}{\sqrt{2}}(1\!\! 1-T_i)$ it
is clear that if the $T_i$ are anti-Hermitian then the $\check{R}_i$
are unitary, giving (c).
\end{proof}

\begin{remark}
Note that if the $T_i$ satisfy $(E1)$ and $(E2)$ then the matrices
$\check{R}_i=\frac{1}{\sqrt{2}}(1\!\! 1+T_i)$ satisfy (\ref{br1})
automatically, and satisfy (\ref{br2}) precisely when
$$(E3^\prime)\quad
T_i+T_iT_{i+1}T_i-T_{i+1}-T_{i+1}T_iT_{i+1}=0.$$ While $(E3)$ is
sufficient to imply $(E3^\prime)$, one wonders if there are other
interesting group relations or almost-complex structures that
satisfy $(E3^\prime)$ but not $(E3)$.
\end{remark}

Let $\phi_{n-1}$ be any representation of $\bE_{n-1}$ satisfying
the hypotheses of Proposition \ref{es2br} and such that
$\phi_{n-1}(e_i)$ is anti-Hermitian.  Then we obtain
unitary representations $\pi_n$ of the braid group $\B_n$ via:
\eq \pi_n ( b_i) = {\frac 1 {\sqrt 2} } (1\!\! 1 +\phi_{n-1} (e_i)
). \en

We wish to describe the images of $\pi_n$ as in \cite{frw}, which is
in fact a special case of our analysis. Note the following key
relations between $\pi_n(\B_n)$ and
$\phi_n(\bE_{n-1})$:
\eqa \label{conj} &&
\pi_n(b_i)\phi_{n-1}(e_{i\pm 1})\pi_n(b_i)^{-1}
=\phi_{n-1} (e_i)\phi_{n-1}(e_{i\pm 1}) , \nonumber\\
 && \pi_n(b_i)\phi_{n-1}(e_j)\pi_n(b_i)^{-1} =\phi_{n-1}(e_j) \ena
 where $1 \le i\pm 1\le n-1$ and $|i-j|\ge 2$. Observing that
$[\pi_n(b_i)]^2=\phi_{n-1}(e_i)$ these relations immediately imply
that the restriction of $\pi_n$ to the normal subgroup $\PP_n$ is
$H_n:=\pi_n(\PP_n)=\phi_{n-1}(\bE_{n-1})$ since $\PP_n$ is
generated by all conjugates of $(b_i)^2$.

Now let us show that the braid group image $G_n:=\pi_n(\B_n)$ is
an extension of $\bE_{n-1}$ by the symmetric group ${\mathcal S}_n$.
We must show that the surjective homomorphism
$\pi^\prime_n:\mathcal{S}_n\to G_n/H_n$ is in fact a bijection.
For this consider the homomorphism $\vartheta:G_n\to \Aut(H_n)$
from $G_n$ to the automorphism group of $H_n$ where $\vartheta(G_n)$
acts by conjugation (explicitly by (\ref{conj})). Note that since
$\vartheta(H_n)=:Inn(H_n)$, the subgroup of inner automorphisms,
the map $\vartheta$ induces a homomorphism
$\overline{\vartheta}: G_n/H_n\to \Out (H_n):=\Aut(H_n)/\Inn(H_n)$.
Clearly $\Inn(H_n)$ acts on $H_n$ by sign changes. To show that
$\pi^\prime_n$ is injective, it is enough to show that
$\overline{\vartheta}\circ\pi^\prime_n:\mathcal{S}_n\to \Out(H_n)$
is injective, i.e. $\ker(\overline{\vartheta}\circ\pi^\prime_n)\lhd S_n$
is trivial.  Since the only nontrivial proper normal subgroup of
$\mathcal{S}_n$ is $A_n$ for $n\geq 5$, it is enough to show that
the images of the permutations $(12)$, $(123)$ and $(12)(34)$ (for
$n=4$) under $\overline{\vartheta}$ are nontrivial--easily
accomplished using (\ref{conj}). Thus we have an exact sequence:
\eq
 1\!\! 1 \rightarrow H_n\rightarrow
  G_n \rightarrow
{\mathcal S}_n\rightarrow 1\!\! 1, \en for all $n\geq 2$, where
$H_n\to G_n$ is inclusion and $G_n\to \mathcal{S}_n$ is induced by
the quotient map and the isomorphism $\pi_n^\prime$.

\subsection{Representations of $\bE_m$ via almost-complex structures}

Now we introduce almost-complex structures and construct representations
of nearly extraspecial 2-groups satisfying the hypotheses of
Proposition \ref{es2br}, which will then give rise to representations
of $\B_n$ generalizing those of \cite{frw}.

First we consider a $2n\times 2n$ anti-Hermitian matrix $M_{2n}$ of the
form \eq \label{acomplex} M_{2n}=\sum_{i=1}^{2n} \epsilon(i)\,
|i\rangle \langle {\bar i} |, \quad {\bar i} =2n +1 -i,
 \en
where the function $\epsilon(i)$ satisfy Eqs. (\ref{step}), and the
Dirac kets $|i\rangle$ form an orthonormal basis.  $M_{2n}$
satisfies $(M_{2n})^2=-1\!\! 1_{2n}$ and $(M_{2n})^\dag=-M_{2n}$.

In what follows, we construct two classes of representations
$\phi_{m}^{(i)}$, $i=1,2$ for the group $\bE_{m}$ in
terms of the almost-complex structure $M_{2n}$.  The first class has
already appeared in \cite{yong4}, but we summarize it here for comparison.

{\bf Class (1)}: The almost-complex structure to be used is a
$(2k)^2\times (2k)^2$ matrix $M^{JJ}$ with complex deformation
parameters $q_{ij} \in \C$, \eq \label{mjj}
 M^{JJ}=\sum_{i,j=-J}^{J} \epsilon(i)
q_{ij}|ij\rangle\langle {\bar i}{\bar j}|, \quad {\bar i}=-i,\,\,
 {\bar j}=-j,\quad J=k-\frac 1 2,\,\, k\in \N
\en where the $\epsilon(i)$ may be arbitrarily
chosen subject to (\ref{step}).
These matrices have already appeared in \cite{yong4} for describing
the GHZ states of an \emph{even} number of objects.  For completeness
we include the following, which is implicit in \cite{yong4}:
\begin{thm}
Define a map $\phi^{(1)}_m$ on the generators of $\bE_m$ in terms
of $M^{JJ}$ by:
 \eq\label{phi3}
 \phi^{(1)}_{m}(e_i) =1\!\! 1_{2k}^{\otimes i-1} \otimes M^{JJ}
 \otimes 1\!\! 1_{2k}^{\otimes m-i},\quad i=1,\cdots, m.
 \en
Then $\phi^{(1)}_m$ defines a $(2k)^{m+1}$-dimensional unitary
representation of $\bE_{m}$ if and only if the parameters $q_{ij}$
in $M^{JJ}$ satisfy the following three constraints:
\eq \label{3constr} q_{ij} q_{{\bar i}\, {\bar j}}=1,
\quad q_{ij} q_{\bar ij}=q_{jl} q_{j \bar l}, \quad q_{ij}^\ast
q_{ij}=1. \en
\end{thm}\label{case1thm}
\begin{proof}  One checks that $T_i:=\phi^{(1)}_m(e_i)$ is anti-Hermitian
and satisfies (E1) and (E3) of Proposition \ref{es2br} if and only if
relations (\ref{3constr}) hold.  Relation (E2) is immediate.
\end{proof}

In the spirit of separation of variables, we assume $q_{ij}=q_{i}
q_j$ and $q_i \neq 0$ to obtain solutions of (\ref{3constr}), \eq
\label{33constr} q_{i} q_{\bar i}=1, \quad q_i^\ast=q_{\bar i},
\quad q_i \in \C \en and recast $M^{JJ}$ into a tensor product of
two matrices $M^\prime_{2k}$ and $P^\prime_{2k}$,
\eq \label{m2kp2k}  M^{JJ}
=M^\prime_{2k}\otimes P^\prime_{2k}, \quad M^\prime_{2k}
=\sum_{i=-J}^J \epsilon(i)q_i|i\rangle\langle \bar i|,\,\,
P^\prime_{2k}=\sum_{j=-J}^J q_j|j\rangle\langle \bar j|. \en
These matrices then satisfy $(M^\prime_{2k})^2=-1\!\! 1_{2k}$
and $(M^\prime_{2k})^\dag =-M^\prime_{2k}$ as well as
\eq
 (P^\prime_{2k})^2 =1\!\! 1_{2k}, \quad (P^\prime_{2k})^\dag
=P^\prime_{2k}, \quad M^\prime_{2k}
 P^\prime_{2k} =-P^\prime_{2k} M^\prime_{2k}.
\en

\begin{remark}
It should be pointed out that from an algebraic point of view
we may assume:
\begin{enumerate}
\item $q_i=1$ for all $i$ and
\item $\epsilon(i)=1$ for $i>0$ and $\epsilon(i)=-1$ for $i<0$.
\end{enumerate}
The reason is that for any choice of $q_i$ and any admissible
choice of signs $\epsilon(i)$ the representations of $\bE_m$
defined via $M^{JJ}$ is unitarily equivalent to those
representations of $\bE_m$ obtained from any other choice of
$q_i$ and $\epsilon(i)$. That is, we may find a global unitary
change of basis (via a diagonal matrix in fact) with respect to
which the matrices $M^\prime_{2k}$ and $P^\prime_{2k}$ may be
assumed to have the form:

\eq
 M_{2k} =\sum_{i=-J}^J \epsilon(i) | i^\prime\rangle \langle
\bar i^\prime|, \quad P_{2k} =\sum_{i=-J}^J | i^\prime\rangle
\langle \bar i^\prime| \en
where $\epsilon(i)$ is as above.

On the other hand, these unimodular deformation parameters can be
understood as phase factors which play key roles in quantum
mechanics, for example, the angle variable $\varphi$ at the
deformation parameter $q_{\frac 1 2 \frac 1 2}$ in $M^{\frac 1 2
\frac 1 2}$ is explained as an angle parameter for the rotation in
the Bloch sphere, see \cite{yong1,yong2}. Moreover, these
unimodular deformation parameters have an interpretation in terms
of Berry phases in quantum mechanics \cite{cxg}. The connection to
the Berry phase is the following: Such phases usually can be
removed by suitable unitary transformations and hence are thought
of as spurious; but under certain conditions in an adiabatic
evolution they give rise to non-trivial boundary effects after
periodic closure. (For the possible relations to boundary issues
(in the context of braiding gates), see also refs.
\cite{aravind,abdm,bb}.)
\end{remark}

{\bf Class (2)}:  There are two natural ways in which to generalize
{\bf Class (1)} in search of representations of $\bE_m$. Firstly, we
may consider more general almost-complex structures of the form
$M_{2k_1}\otimes P_{2k_2}$ where $k_1\neq k_2$. Secondly, we look for
solutions to the generalized YBE that satisfy both (\ref{br1}) and
(\ref{br2}). Taken in tandem, this is a formidable problem.  For
simplicity and with an eye towards GHZ states, we consider a special
case: $k_1=1$ and $k_2=2^{N-2}$, $N\ge 2$.  In particular we define
$$M_{2^N}= M_{2}\otimes P_{2^{N-1}} =\sqrt{-1} \sigma_y\otimes
\sigma_x^{\otimes N-1}.$$ Notice that this $M_{2^N}$ does not depend
on any deformation parameters or sign choices $\epsilon(i)$.
Having fixed this $M_{2^N}$ we can construct another class of
representations of $\bE_m$, as summarized by the following main result:

\begin{thm}\label{dimthm} Define
$\phi^{(2)}_m$ on generators of $\bE_m$ by
 \eq\label{phi2}
 \phi^{(2)}_{m}(e_i) =1\!\! 1_{2^k}^{\otimes i-1} \otimes
\sqrt{-1} \sigma_y \otimes \sigma_x^{\otimes N-1}
 \otimes 1\!\! 1_{2^k}^{\otimes m-i},\quad i=1,\cdots, m.
 \en
 Then $\phi^{(2)}_m$
defines an (anti-Hermitian) representation of
$\bE_m$ into ${\rm
U}(2^{N+k(m-1)})$ for all $m\geq 2$ if and only if
$\frac N 2 \le k \le N-1$.
\end{thm}
\begin{proof}
One easily checks that $M_{2^N}=\sqrt{-1} \sigma_y \otimes
\sigma_x^{\otimes N-1}$ is anti-Hermitian and satisfied
$M_{2^N}^2=-\1_{2^N}$.  This implies that the
$\phi^{(2)}_{m}(e_i)$ are also anti-Hermitian and satisfy
$$[\phi^{(2)}_{m}(e_i)]^2=-1\!\! 1_{2^{N+k(m-1)}}.$$ Direct
calculation shows that $\phi^{(2)}_{m}(e_i)$ and
$\phi^{(2)}_{m}(e_j)$ with $i\le j$ anti-commute if and only if
$1\le k(j-i)\le N-1$, and commute if and only if $N-1 < k(j-i)$.
Taking $j=i+1$ in the first condition yields $1\le k \le N-1$, while
for $m\ge 3$ taking $j=i+2$ in the second condition produces $\frac
N 2 \le k \le N-1$.

\end{proof}

We remark that while $\phi_m^{(2)}$ is a 2-parameter family of
representations depending on $N$ and $k$, we have suppressed this
dependence for notational convenience.

\subsection{Decomposition into irreducible representations of $\B_n$}

We proceed to determine the decomposition of the unitary braid group
representations obtained from the representations
$\phi^{(i)}_{n-1}$ ($i=1,2$) of $\bE_{n-1}$ into irreducible
constituents.  Since these $\phi^{(i)}$ satisfy the hypotheses of
Proposition \ref{es2br}, we have the conclusions at our disposal.
Moreover, the classification and formulas for irreducible representations
$\rho_1$, $\lambda_1$ and $\lambda_2$  of group $\bE_{n-1}$ are given
in \cite{frw}, so one can easily compute their (irreducible, since the
restrictions to $\PP_n$ are) extensions $\hat{\rho}_1$, $\hat{\lambda}_1$
and $\hat{\lambda}_2$ to $\B_n$.

We consider {\bf Class (1)} and {\bf Class (2)} simultaneously, so
that the representation $\pi_n$ is induced from $\phi^{(i)}_{n-1}$
with $i=1$ or $i=2$. First let us
consider the decomposition of $\pi_n$ of $\B_n$ with $n$ odd, so
that the restriction of $\pi_n$ to $\PP_n$ factors over the
representation $\phi^{(i)}_{n-1}$ of $\bE_{n-1}$ with $n-1$ even.
Set $d^o_i=\frac{\dim(\phi_{n-1}^{(i)})}{2^{(n-1)/2}}$. Then Proposition
\ref{es2br}(d) implies that $\phi^{(i)}_{n-1}$ decomposes into
irreducible subrepresentations as $d^o_i$ copies of the
$2^{(n-1)/2}$-dimensional irreducible $\bE_{n-1}$-representation
$(\rho_1,V_1)$.  Thus $\pi_n$ decomposes as $d^o_i$ copies of
$\hat{\rho}_1$ as representations of $\B_n$.  Now consider $n$ even,
and set $d^e_i=\frac{\dim(\phi^{(i)}_{n-1})}{2^{n/2}}$. It was
observed in \cite{frw} that the restriction of $\rho_1$ to $\bE_{n-1}$
decomposes as follows: $${\rm Res}^{\bE_{n}}_{\bE_{n-1}}(\rho_1,V_1)
=(\lambda_1,W_1)\oplus(\lambda_2,W_2).$$
 This together with Proposition
\ref{es2br}(a) shows that $\phi^{(i)}_{n-1}$ decomposes as a
representation of $\bE_{n-1}$ into $d_i^e$ copies of
$\lambda_1\oplus\lambda_2$ where $\lambda_i$, $i=1,2$ are the two
inequivalent irreducible representations of $\bE_{n-1}$ with $n$
even with $\dim(\lambda_i)=2^{(n-2)/2}$. Thus the $\B_n$
representation $\pi_n$ with $n$ even decomposes as $d^e_i$ copies of
$\hat{\lambda}_1 \oplus\hat{\lambda}_2$. For completeness, let us
recall that $\dim(\phi^{(1)}_{n-1})=(2k)^n$ and
$\dim(\phi^{(2)}_{n-1})=2^{N+k(n-2)}$ so that $d^o_1=k^n2^{(n+1)/2}$
and $d^o_2=2^{N+k(n-2)-(n-1)/2}$ with $d^e_i$ computed similarly.

In \cite{frw} it was shown that a renormalization of the $4\times 4$
Bell basis-change matrix leads to a (projectively equivalent)
representation of $\B_n$ that factors over the well-known Jones
representation at a $4$th root of unity (via Temperley-Lieb algebras).
It is clear that the generalized Bell matrices $B_{2k}$ (with the case
in \cite{frw} as a special case) may be renormalized in the same fashion
to obtain the same conclusion.  In particular, the matrices $B_{2k}$ may
be used to define link-invariants, which will contain the same topological
information as the Jones polynomial \cite{jones2,jones3} at a $4$th root
of unity.

\section{GHZ states and their unitary evolution via QYBE}

In \cite{yong4}, it was observed that GHZ states (corresponding to an
even number of qubits) can be obtained from the product basis via
operators that satisfy the (conventional) YBE (\ref{ybe}).  In what
follows we put the results of \cite{yong4} into the present context
and describe the role played by the generalized YBE (\ref{gybe}) in
producing GHZ states with an \emph{odd} number of qubits. Moreover,
Yang--Baxterization \cite{jones1,gxw} is exploited to obtain specific
Hamiltonians that give rise to unitary evolution of the GHZ states.

\subsection{Unitary basis transformation matrices}

The two dimensional Hilbert space ${\cal H}_2$ spanned by
eigenvectors $|m\rangle, m=\pm \frac 1 2$ of the spin-$\frac 1 2$
operators (i.e. Pauli matrices, for example, $\sigma_z|\pm
\frac 1 2\rangle=\pm |\pm\frac 1 2\rangle$), has the following
realization of coordinate vectors over the complex field $\C^2$, \eq
|\frac 1 2\rangle :=\left(\begin{array}{c} 1 \\ 0
\end{array}\right), \quad  |-\frac 1 2\rangle
:=\left(\begin{array}{c} 0 \\ 1 \end{array}\right), \quad \alpha
 |\frac 1 2\rangle + \beta |-\frac 1 2\rangle
=\left(\begin{array}{c} \alpha \\
\beta \end{array}\right),\en which determine actions of
$\sqrt{-1}\sigma_y$ and $\sigma_x$ on the basis $|m\rangle$, \eq
 \label{con_step}
\sigma_x |m\rangle=|\bar m\rangle, \quad \sqrt{-1}\sigma_y|m\rangle
=\epsilon^\prime(m)|\bar m\rangle, \quad \bar m =-m, \quad m=\pm
\frac 1 2 \en where $\epsilon^\prime(\frac 1
2)=-\epsilon^\prime(-\frac 1 2)=-1$. A state vector in this ${\cal
H}_2$ is usually called a qubit in quantum information theory
\cite{nielson}.

The Hilbert space ${\cal H}_{2^N}$ is isomorphic to
$(\C^2)^{\otimes N}$ and describes a physical system consisting of
$N$ qubits, each qubit with two linearly independent states. It has
an orthonormal basis denoted by Dirac kets $|\Phi_k\rangle$,
$1\le k \le 2^{N}$ which are tensor products of
$|\pm \frac 1 2\rangle$ by
\eq
 |\Phi_k\rangle \equiv |m_1,\cdots,m_N\rangle \equiv|m_1\rangle\otimes
  \cdots \otimes |m_N\rangle,
 \quad m_1, \cdots, m_N =\pm \frac 1 2.
   \en
Here the lower index $k$ is a given function of $m_1,\cdots m_N$,
see \cite{fujii},  \eq \label{convention}
k[m_1,\cdots,m_N]=2^{N-1}+\frac 1 2-\sum_{i=1}^N 2^{N-i}\,\, m_i,
 \en
so that in coordinates
  \eq \label{phik} |\Phi_{k}\rangle
 =(\underbrace{0,\cdots, 0}_{k-1}, 1, \underbrace{0,\cdots,
 0}_{2^{N}-k})^T,
 \en
where ${}^T$ denotes transpose.

This orthonormal basis $|\Phi_k\rangle$ is partitioned into two
sets respectively denoted by Dirac kets $|\Phi_l\rangle$ and
$|\Phi_{\bar l}\rangle$, $1\le l \le 2^{N-1}$ and $\bar l=2^N-l+1$,
 \eq
 |\Phi_l\rangle=|m_1,\cdots,m_N\rangle, \quad |\Phi_{\bar l}\rangle
  =|\bar m_1,\cdots,\bar m_N\rangle, \quad \bar m_i=-m_i,\,\, 1\le l \le
  N. \en
In terms of $|\Phi_l\rangle$ and $|\Phi_{\bar l}\rangle$, the
Hilbert space ${\cal H}_{2^N}$ is spanned by the $2^N$ orthonormal
GHZ states $|\Psi_l\rangle$ of $N$ qubits, \eq \label{ghz}
 |\Psi_l\rangle \equiv \frac  1 {\sqrt 2} (|\Phi_l\rangle + |\Phi_{\bar l}\rangle),
 \quad  |\Psi_{\bar l}\rangle \equiv\frac 1 {\sqrt 2} (|\Phi_l\rangle -
|\Phi_{\bar l}\rangle ).
 \en
These GHZ states are maximally entangled states that have
been widely used in quantum information theory
\cite{ghz1,ghz2,ghz3}. The set of all GHZ states forms an
orthonormal basis of ${\cal H}_{2^N}$.

As an example, consider the Hilbert space $\C^2\otimes \C^2$ for two
qubits (here GHZ states are the well-known Bell states). In terms of
the orthonormal product basis $|\Phi_k\rangle$, $k=1,\cdots,
4$,
 \eqa && |\Phi_1\rangle=
|\frac 1 2,\frac 1 2\rangle, \quad |\Phi_4\rangle=|\Phi_{\bar
1}\rangle= |-\frac 1 2, -\frac 1
2\rangle, \nonumber\\
& & |\Phi_2\rangle= |\frac 1 2, -\frac 1 2\rangle, \quad
|\Phi_3\rangle=|\Phi_{\bar 2}\rangle= |-\frac 1 2,\frac 1 2\rangle,
 \ena
where the numbering for lower indices is consistent with the
convention (\ref{convention}), \eq k[\frac 1 2,\frac 1 2]=1,\,\,
k[\frac 1 2,-\frac 1 2]=2,\,\, k[-\frac 1 2,\frac 1
2]=3,\,\,k[-\frac 1 2,-\frac 1 2]=4, \en Bell states have the same
formulations as their conventions,
 \eqa
 \label{bell}
&& |\Psi_1\rangle=\frac 1 {\sqrt 2} (|\Phi_1\rangle + |\Phi_{\bar
1}\rangle), \quad |\Psi_4\rangle=|\Psi_{\bar 1}\rangle=\frac 1
{\sqrt 2} (|\Phi_1\rangle - |\Phi_{\bar 1}\rangle),
\nonumber\\
&& |\Psi_2\rangle=\frac 1 {\sqrt 2} (|\Phi_2\rangle + |\Phi_{\bar
2}\rangle), \quad |\Psi_3\rangle=|\Psi_{\bar 2}\rangle=\frac 1
{\sqrt 2} (|\Phi_2\rangle - |\Phi_{\bar 2}\rangle).
 \ena

\subsection{Unitary braid representations for GHZ states}

It is not difficult (mathematically) to construct unitary
operators $U$ on ${\cal H}_{2^N}$ that generate the GHZ states
$|\Psi_j\rangle$ from the product basis $|\Phi_k\rangle$. In
quantum circuits, the gate operators are physically realized as
unitary evolutions of some system. For braiding gates, we
particularly want the evolution operator $U$ to satisfy the
generalized Yang-Baxter equations (\ref{gybe}), in order to give
rise to representations of the braid group.

\subsubsection{An even number of qubits}

We recapitulate the results of \cite{yong4} for GHZ states of an
{\it even} number $2n$ of qubits. These are associated to the
(generalized) Bell matrix $B^{JJ}$ in terms of the almost-complex
structure $M^{JJ}$ in {\bf Class(1)},
\eq
 B^{JJ} =1\!\! 1_{(2k)^2} + M^{JJ}, \quad J=k-\frac 1 2,\quad k\in \N.
\en
An important point is that $B^{JJ}$ is a $(2k)^2\times (2k)^2$
matrix, while the dimension of the Hilbert space spanned by the GHZ
states of $2n$ qubits is $2^{2n}$. So for the GHZ states generated by
$B^{JJ}$ to {\it span} a $2^{2n}$-dimensional Hilbert space, one needs
$J=2^{n -1}-\frac 1 2$.

Thus for {\bf Class (1)} with unimodular deformation
 parameters are chosen to be $1$,
which give rise to the unitary braid representation
$B^{JJ}_{2^{2n}}$:
 \eq
  B^{JJ}_{2^{2n}}= \frac 1 {\sqrt 2} (1\!\! 1_{2^{2n}} + M^{JJ}_{2^{2n}}),
  \quad M^{JJ}_{2^{2n}}=\sqrt{-1}\sigma_y\otimes \sigma_x^{\otimes 2n-1}.
 \en
The GHZ states of $2n$ qubits obtained by the corresponding Bell matrix
on the orthonormal product basis $|\Phi_k\rangle$ with $\epsilon^\prime(m_1)$
as in (\ref{con_step}) are
 \eq
  \label{2nqubit}
 \frac 1 {\sqrt 2} (|m_1,\cdots, m_{2n}\rangle
  + \epsilon^\prime(m_1) |\bar m_1, \cdots, \bar m_{2n}\rangle)
 \en
which leads to the  unitary basis transformation matrix,
 \eq
 B^{JJ}_{2^{2n}}=(|\Psi_{2^{2n}}\rangle,
|\Psi_{2^{2n}-1}\rangle,  \cdots,
 |\Psi_{2}\rangle, |\Psi_{1}\rangle).
 \en

\subsubsection{An odd number of qubits}

Obviously, unitary braid representations in {\bf Class (1)} can not
yield the GHZ states of an \emph{odd} number, say $2n+1$, of qubits.
But the unitary braid representations in {\bf Class (2)} can,
via the matrices \eq B_{2^{2n+1}}=\frac 1 {\sqrt 2} (1\!\!
1_{2^{2n+1}} +M_{2^{2n+1}}),\quad
 M_{2^{2n+1}}= {\sqrt -1}\sigma_y \otimes (\sigma_x)^{\otimes 2n},
\en
The essential differences between unitary braid representations
in {\bf Class (1) } and {\bf Class (2)} are: every strand for
the braid group  in {\bf Class (1)} lives in the same dimensional
vector space, whereas this is not always true in {\bf Class (2)},
\emph{e.g.}, for the Bell matrix $B_{2^{2n+1}}$.

The GHZ states of $2n+1$ qubits generated by the Bell matrix
$B_{2^{2n+1}}$ in {\bf Class (2)} acting on the product basis
$|\Phi_l\rangle$ and $|\Phi_{\bar l}\rangle$ have a similar form
as shown in (\ref{2nqubit}), and $B_{2^{2n+1}}$ represents the
unitary basis transformation matrix by
 \eq
 B_{2^{2n+1}}=(|\Psi_{2^{2n+1}}\rangle, |\Psi_{2^{2n}}\rangle,  \cdots,
 |\Psi_{2}\rangle, |\Psi_{1}\rangle).
 \en
For example, on the product basis $|\Phi_l\rangle$ for three qubits
the generalized Bell matrix obtained from the almost complex structure
$M_{8}=\sqrt{-1}\sigma_y\otimes \sigma_x^{\otimes 2}$ takes the form
\eq
B_{8} =\frac 1 {\sqrt
2}\left(\begin{array}{llllllll}
 1 & 0 & 0 & 0 & 0 & 0 & 0 & 1 \\
  0 & 1 & 0 & 0 & 0 & 0 & 1 & 0 \\
  0 & 0 & 1 & 0 & 0 & 1 & 0 & 0 \\
   0 & 0 & 0 & 1 & 1 & 0 & 0 & 0 \\
  0 & 0 & 0 & -1 & 1 & 0 & 0 & 0 \\
  0 & 0 & -1 & 0 & 0 & 1 & 0 & 0 \\
  0 & -1 & 0 & 0 & 0 & 0 & 1 & 0 \\
   -1 & 0 & 0 & 0 & 0 & 0 & 0 & 1 \\
\end{array} \right)
  \en
which produces all the GHZ states $|\Psi_l\rangle$ of three qubits
\cite{ghz1,ghz2,ghz3} and satisfies the conditions of Theorem \ref{dimthm}
with $k=2$.  Thus for each $m\geq 1$ one obtains 1) $2^{2m+1}$-dimensional
representations of $\B_m$ so that 2) the action of the braid generators on
the product basis produces (higher dimensional promotions of) GHZ states
on three qubits.

As a basis change operator $B_{8}$ takes the basis $|\Phi_l\rangle$
(with the usual ordering) to
\eq
(|\Psi_{\bar 1}\rangle, |\Psi_{\bar 2}\rangle,  |\Psi_{\bar 3}\rangle,
 |\Psi_{\bar 4}\rangle, |\Psi_4\rangle, |\Psi_3\rangle,
 |\Psi_{2}\rangle, |\Psi_{1}\rangle).
\en

\subsection{Unitary evolution of GHZ states}

Unitary evolution of GHZ states as well as the corresponding
Sch{\"o}dinger equation can be explored with the help of
Yang--Baxterization \cite{jones1,gxw}, and this is a systematic
elaboration of previous research work.  Unitary evolution of Bell
states have been discussed in detail \cite{yong1,yong2}, while
unitary evolution of GHZ states have been only briefly sketched in
\cite{yong4}.

The quantum Yang--Baxter equation (QYBE) is of the form \eq
\label{qybe}
 \check{R}_i(x)\,\check{R}_{i+1}(x y)\,\check{R}_i(y)=
 \check{R}_{i+1}(y)\,\check{R}_i(x y)\,\check{R}_{i+1}(x)\en with
 $x$ or $y$ the spectral parameter. It is well known that one
 can set up an integrable model by following a given recipe in
 terms of a solution of the QYBE, see \cite{yang,baxter}. At $x=y=0$,
 obviously, the QYBE reduces to $\check{R}_i\check{R}_{i+1}\check{R}_i
 =\check{R}_{i+1}\check{R}_i\check{R}_{i+1}$, the same as the braid
 group relation (\ref{br2}), $b_ib_{i+1}b_i=b_{i+1}b_ib_{i+1}$.
 Hence a solution $\check{R}(x)$ of the QYBE always reduces to a
 braid group representation $b=\check{R}(0)$. In other words, $\check{R}(0)=b$
 can be regarded as the asymptotic condition of a solution $\check{R}(x)$
 of the QYBE. Conversely, similar to a procedure of solving a differential
 equation with specified initial-boundary conditions, Baxterization  \cite{jones1}
 or Yang--Baxterization \cite{gxw}, represents a procedure of
 constructing a solution $\check{R}(x)$ of the QYBE (\ref{qybe}) with
 the asymptotic condition, $\check{R}(0)=b$, where the braiding $b$-matrix
 has been specified. For example, for a $b$-matrix with two distinct
 eigenvalues $\lambda_1$ and $\lambda_2$, the corresponding $\check{R}(x)$-matrix
 obtained with Yang--Baxterization is found to be of the form
 \eq
 \check{R}(x)=b +x \lambda_1\lambda_2 b^{-1}.
 \en
 Please refer to Appendix A of the paper \cite{yong2} for
 more details.

The unitary braid operator $B$ derived from the
almost-complex structure $M$ (\ref{acomplex}), has two distinct
eigenvalues $\zeta$ and $\zeta^\ast$ and satisfies
\eq  \label{characteristic} (B-\zeta\,\, 1\!\!
1) (B- \zeta^\ast\,\, 1 \!\! 1)=0.
 \en  Using
Yang--Baxterization, a solution of the QYBE (\ref{qybe}) with the
asymptotic limit $B$, is
 \eq
 \label{rx}
 \check{R}(x)=B+x B^{-1}=\frac 1 {\sqrt 2} (1+x)1\!\! 1
  +\frac 1 {\sqrt 2}(1-x)M.
 \en
where the lower indices of $B, M, 1\!\! 1$ are suppressed for
convenience. This
$\check{R}(x)$-matrix can be updated to be a unitary matrix $B(x)$
by adding a normalization factor $\rho(x)$, i.e.,
 \eq
B(x)=\rho^{-\frac 1 2}\check{R}(x), \qquad \rho= 1+x^2,\,\,
  x\in {\mathbb R}.
 \en

As the real spectral parameter $x$  plays the role of the time
variable, the Schr{\"o}dinger equation describing the unitary
evolution of a state $\psi(0)$ (independent of $x$) determined by
the $B(x)$ matrix, i.e., $\psi(x)=B(x)\psi(0)$, has the form
 \eq
 \label{scr1}
 \sqrt{-1}\frac {\partial} {\partial x}\psi(x)=H(x)\psi(x),\qquad
  H(x)\equiv \sqrt{-1}\frac {\partial B(x)} {\partial x}  B^{-1}(x),
 \en
where the time-dependent Hamiltonian $H(x)$ is  given by
 \eq
H(x)= \sqrt{-1} \frac {\partial } {\partial x}(\rho^{-\frac 1 2}
{\check R}(x))(\rho^{-\frac 1 2}
 \check{R}(x))^{-1}=- \sqrt{-1}  \rho^{-1} M.
\en To construct the time-independent Hamiltonian, the spectral parameter $x$ is
replaced by a new time
variable $\theta$ by the change of variables \eq \cos\theta=\frac 1 {\sqrt{1+x^2}},
   \qquad
 \sin\theta=\frac x {\sqrt{1+x^2}},
 \en
so that the unitary matrix $B(x)$ has a new formulation in terms of
$\theta$,
 \eq
 B(\theta)=\cos\theta B+\sin\theta B^{-1}
  =e^{(\frac \pi 4-\theta) M},
  \en
and hence the Schr{\"o}dinger equation for the time evolution of
$\psi(\theta)=B(\theta)\psi(0)$ is given by
 \eq
 \label{scr2}
 \sqrt{-1}\frac {\partial} {\partial
 \theta}\psi(\theta)=H\psi(\theta),\qquad
 H\equiv\sqrt{-1}\frac {\partial B(\theta)}
 {\partial \theta}  B^{-1}(\theta)=-\sqrt{-1} M,
 \en
where the time-independent Hamiltonian  $H$ is Hermitian since the
almost-complex structure $M$ is anti-Hermitian. The unitary
time-evolution operator $U(\theta)$ has the form
$U(\theta)=e^{-\theta M}$. Furthermore, with the shifted time
variable $\theta^\prime$, unitary matrices $B(\theta^\prime)$ and
$U(\theta^\prime)$ take the same form, \eq
 B(\theta^\prime)=U(\theta^\prime) = e^{-\theta^\prime M}, \quad
 \theta^\prime = \theta -\frac \pi 4.
\en

In {\bf Class (1) }, Yang--Baxterization of the Bell matrix $B^{JJ}$
for GHZ states of $2n$ qubits has the form
 \eq
 B^{JJ}(\theta^\prime) =\cos\theta^\prime 1\!\! 1_{2^{2n}}
 -\sin\theta^\prime M_{2^n} \otimes P_{2^n}.
 \en
Define
$|\alpha\rangle$ by

\eq
 \label{alpha}
 |\mu\rangle:=|\Phi_\mu\rangle, \quad
 |\nu\rangle:=|\Phi_\nu\rangle, \quad
 |\alpha\rangle:=|\mu\nu\rangle=|\Phi_{(\mu-1) 2^n +\nu}\rangle.
 \en
Then
 the unitary evolution of the GHZ state $|\alpha\rangle$
is given by
 \eq
 B^{JJ}(\theta^\prime)|\alpha\rangle
  =\cos\theta^\prime |\alpha\rangle -\sin\theta^\prime \epsilon(\mu)
  e^{\frac {\varphi_\mu +\varphi_v} 2} |\bar \alpha \rangle
 \en
corresponding to the Hamiltonian $H^{JJ}=-\sqrt{-1} M_{2^n}\otimes
P_{2^n}$.

In {\bf Class (2)}, Yang--Baxterization of the Bell matrix $B_{2^n}$
has the form in terms of the Hamiltonian $H_{2^n}$,
 \eq
B_{2^n}(\theta^\prime)=e^{-\sqrt{-1} \theta^\prime H_{2^n}},
\qquad H_{2^n}
 =\sigma_y \otimes \sigma_x^{\otimes n-1}
 \en
which derives the unitary evolution of GHZ state $|\Psi_l\rangle$
defined by (\ref{ghz}),
 \eq
B_{2^n}(\theta^\prime) |\Phi_l\rangle
 =\cos\theta^\prime |\Phi_l\rangle - \sin\theta^\prime
 \epsilon^\prime(m_1) |\Phi_{\bar l}\rangle.
 \en
For example, the unitary evolution of GHZ states of three qubits
 determined by $B_{8}(\theta^\prime)=e^{-\sqrt{-1}\theta^\prime H_{8}}$
 have the following realization,
  \eq
  B_{8}(\theta^\prime)|\frac 1 2 \frac 1 2 \frac 1 2\rangle
  =\cos\theta^\prime |\frac 1 2\frac 1 2\frac 1 2\rangle
  +\sin\theta^\prime |\frac {-1} 2 \frac {-1} 2 \frac {-1} 2
  \rangle, \quad H_{8}=\sigma_y\otimes \sigma_x^{\otimes 2}.
  \en

\section{Conclusions and Discussions}

In this paper we have revealed the connections between a special class
of multi-qubit braiding quantum gates and extraspecial 2-groups. More
concretely, we have shown that one may associate to certain almost-complex
structures anti-Hermitian representations of extraspecial 2-groups,
which in turn give rise to unitary representations of the braid
group factoring over extensions of 2-groups by symmetric groups.
These unitary braid representations can be used to generate the
maximally entangled GHZ states for an arbitrary number of qubits.
Since the braiding quantum gates are known to play a pivotal role in
topological quantum computation \cite{kauffman1,kitaev,fklw,kl}, our
present work suggests that extraspecial 2-groups should play an important
role in topological quantum computation, at least in the analysis of
quantum circuits consisting of the braiding gates that we have studied.

For example, in the Freedman-Kitaev topological model \cite{kitaev,fklw}
for quantum computation, the gates are realized as operators representing
braids in $2+1$ dimensions. Given a gate $U$ it is a difficult problem
to find a braid that (even approximately) realizes $U$.  Indeed, the
ubiquitous entangling CNOT gate is quite difficult to achieve in this
setting (see e.g. \cite{bonetal}). It
is not known if the CNOT gate can be \emph{exactly} realized in some
model. Our results by exploring extraspecial 2-groups suggest that in
the $SU(2)_2$ model for the topological state (see \cite{DSetal})
realized as the state at filling fraction $\nu=5/2$ in the fractional
quantum Hall effect, one may obtain all GHZ states \emph{exactly}.
In this sense the generalized Bell matrices we have studied here are
as important to topological quantum computation as the CNOT gate is
to the quantum circuit model.

We also expect that extraspecial 2-groups play an important role
in the theory of quantum error correction, which protects quantum
information against noises. On one hand, the extraspecial 2-groups
provide a bridge between quantum error correcting codes and binary
orthogonal geometry \cite{crss1}. On the other hand, they form a
subgroup of the Pauli group \cite{crss2}, which plays a crucial
role in the theory of stabilizer codes \cite{gottesman}. Therefore,
the new connection, that we have revealed in this paper, between
extraspecial 2-groups and braid group representations suggests
possible applications of the multi-qubit unitary braiding quantum
gates in quantum error correction codes. In particular, the Jones braid
representations at a 4-th root of unity is known to be closely related
to the representations of extraspecial 2-groups \cite{frw}. While the
finiteness of the braid group image precludes the associated braiding
gates alone from forming a universal gate set (in the sense of
\cite{flw}), this new connection suggests in turn that quantum systems
with braiding statistics modeled by the Jones representation at 4-th
roots of unity may be used for quantum error correction.

Finally we conclude with a mathematical remark. Goldschmidt and
Jones \cite{GJ} use extraspecial $p$-groups (Heisenberg groups) to
construct braid group (specialized Burau-Squier) representations
factoring over finite symplectic groups. Although they work
exclusively over fields of odd characteristic and hence only
extraspecial $p$-groups with $p\neq 2$ appear, our work may be
thought of as a characteristic-2 complement to \cite{GJ} in some
respects, particularly as link invariants are concerned.

 \section*{Acknowledgments}

We thank Zhenghan Wang for participating in the early stage
of this work. Y. Zhang thanks L.H. Kauffman for stimulating
discussions and S.L. Braunstein for helpful comments. E.
Rowell was supported in part by NSA grant H98230-08-1-0020.
Y. Zhang was supported in part by the Seed Grant of University
of Utah and NSFC--10605035. Y.S. Wu was supported in part by
U.S. NSF through Grant No. PHY-0407187.

{\em Note Added:} Deformations with phase parameters similar to
those appearing in eq. (3.12) have been used in the context of
deformed exchange statistics in 1D systems, leading to a deformed
solution of the Yang-Baxter equation \cite{oster2000}. Also the
matrix (3.11) recently appears in the context of an SL(2)
invariant extension of the entanglement measure concurrence to
higher (half-integer) spins in ArXiv. 0908.3818
(\cite{oster2009}). We thank the referee for pointing out these
references to us.

\appendix

\section{The generalized quantum Yang--Baxter equation}

The $\check{R}_i(x)$-matrix in terms of the representation
$\phi_{n-1}(e_i)$ of the extraspecial 2-group ${\bE}_n$ of the
form
 \eq
 \check{R}_i(x) = \frac 1 {\sqrt 2} ( (1+x) 1\!\! 1 +(1-x)
\phi_{n-1}(e_i) )
 \en
can be shown to satisfy the QYBE (\ref{qybe}). The proof is
by calculation:  \eqa
 && \sqrt 2\check{R}_i(x)\check{R}_{i+1}(x y)\check{R}_i(y)= (
   (1+x y) (x+y) 1\!\! 1 \nonumber\\
 &&  +(1-x y)(y-x) \phi_{n-1}(e_i) \phi_{n-1}(e_{i+1})
   \nonumber\\
 &&  + (1+x y) (1-x y) (\phi_{n-1}(e_i) +\phi_{n-1}(e_{i+1}))
   ) \ena
which is symmetric under  $i\leftrightarrow i+1 $ and $x
\leftrightarrow y$ so that $\check{R}_i(x)$ satisfies (\ref{qybe}).

In the {\bf Class (1)} representation for ${\bE}_n$, equation
(\ref{qybe}) is the standard version of the QYBE exploited in the
literature, i.e.,
$$ (\check{R}(x)\otimes 1\!\! 1_{2k})(1\!\! 1_{2 k}\otimes
\check{R}(x y))(\check{R}(y)\otimes 1\!\! 1_{2k})= (1\!\! 1_{2k}
\otimes \check{R}(y)) (\check{R}(x y)\otimes 1\!\! 1_{2k}) (1\!\!
1_{2k}\otimes \check{R}(x))
 $$
where $\check{R}_i(x)$ is an invertible $2k\otimes 2 k$ matrix.
However in the {\bf Class (2)} representation for ${\bE}_n$,  equation
(\ref{qybe}) is an example of a {\em generalized version} of the
QYBE given by $$ (\check{R}(x)\otimes 1\!\! 1_{l})(1\!\!
1_{l}\otimes \check{R}(x y))(\check{R}(y)\otimes 1\!\! 1_{l})=
(1\!\! 1_{l} \otimes \check{R}(y)) (\check{R}(x y)\otimes 1\!\!
1_{l}) (1\!\! 1_{l}\otimes \check{R}(x))
 $$
 where $l=p^k$ ($2\le p \in \N$) and $\check{R}(x)$ is an invertible
 $p^N\otimes p^N$ matrix, in other words, (\ref{qybe}) is a
Yang--Baxterized version of the generalized YBE (\ref{gybe}).

By reparameterization, the above $\check{R}(x)$-matrix has an
alternative expression, $\check{R}_i(\Theta)$, given by
  \eq
 \check{R}_i(\Theta) =1 \!\! 1 + \tanh \Theta\,\, \phi_{n-1}(e_i)
  \en
satisfying the parameter-additive (instead of multiplicative) form
of the QYBE (\ref{qybe}),
\eq
 \check{R}_i(\Theta_1)\check{R}_{i+1}(\Theta_1 +\Theta_2)
 \check{R}_i(\Theta_2) =\check{R}_{i+1}(\Theta_2)
\check{R}_{i}(\Theta_1 +\Theta_2)
 \check{R}_{i+1}(\Theta_1).
\en
This form is more closely related to elastic or inelastic scattering
phenomena in quantum physics.  Interested readers are invited to refer
to \cite{yong3,yong4}.


\begin{thebibliography}{99}

 \bibitem{dye}
  H.A. Dye, {\it Unitary Solutions to the Yang--Baxter Equation in Dimension Four},
  Quant. Inf. Proc. {\bf 2} (2003) 117-150. Arxiv:  quant-ph/0211050.

  \bibitem{kauffman1}
 L.H. Kauffman and S.J. Lomonaco Jr.,  {\it Braiding Operators are
 Universal Quantum Gates},  New J. Phys. {\bf 6} (2004) 134.
 Arxiv: quant-ph/0401090.

 \bibitem{yong1} Y. Zhang, L.H. Kauffman and M.L. Ge,
 {\it Universal Quantum Gate, Yang--Baxterization and Hamiltonian}.
  Int. J. Quant. Inform. {\bf  3} no. 4 (2005) 669-678. Arxiv: quant-ph/0412095.

 \bibitem{yong2} Y. Zhang, L.H. Kauffman and M.L. Ge,
 {\it Yang--Baxterizations, Universal Quantum Gates and
  Hamiltonians}.  Quant. Inf. Proc. {\bf 4} (2005) 159-197. Arxiv: quant-ph/0502015.

  \bibitem{frw} J. Franko, E.C. Rowell and Z. Wang,
 {\it Extraspecial 2-Groups and Images of Braid Group
 Representations}. J. Knot Theory Ramifications, {\bf 15} (2006) 413-428.
 Arxiv: math.RT/0503435.

\bibitem{zkw} Y. Zhang, L.H. Kauffman and R.F. Werner, {\it Permutation and
its Partial Transpose}.   Int. J. Quant. Inform. {\bf 5} no. 4 (2006) 469-507.
Arxiv: quant-ph/0606005.

\bibitem{yong3} Y. Zhang, N. Jing and M.L. Ge,
  {\it Quantum Algebras Associated With Bell States}.  J. Phys. A: Math.
  Theor. {\bf 41} (2008) 055310; Arxiv: math-ph/0610036.

\bibitem{yong4} Y. Zhang and M.L. Ge,
 {\it GHZ States, Almost-Complex Structure and Yang--Baxter Equation}.
 Quant. Inf. Proc. {\bf 6} (2007) 363-379; Arxiv: quant-ph/0701244.

\bibitem{yang} C.N. Yang,
 {\it Some Exact Results for the Many Body Problems in One
 Dimension with Repulsive Delta Function Interaction},
 Phys. Rev. Lett. {\bf 19} (1967) 1312-1314.

 \bibitem{baxter} R.J. Baxter, {\it Partition Function of the
 Eight-Vertex Lattice Model}, Annals Phys.\ {\bf 70} (1972) 193-228.



 \bibitem{ghz1} D.M. Greenberger, M.A. Horne and A. Zeilinger,
 {\it Going beyond Bell's Theorem}, in {\em Bell's Theorem, Quantum
 Theory, and Conceptions of the Universe}, edited by M. Kafatos, pp.
 73--76, (Kluwer Academic, Dordrecht, 1989).

\bibitem{jones1} V.F.R. Jones, {\it Baxterization}, Int.\ J.\ Mod.\ Phys. {\bf A6}
(1991) 2035-2043.

\bibitem{gxw} M.L. Ge, K. Xue and Y-S. Wu, {\it Explicit Trigonometric
Yang--Baxterization}, Int.\ J.\ Mod.\ Phys. {\bf A6} (1991) 3735.





 \bibitem{ghz2} D.M. Greenberger, M.A. Horne, A. Shimony, and A.Zeilinger,
    {\it Bell's Theorem Without Inequalities}, Am. J. Phys. {\bf 58} (1990) 1131-1143.

 \bibitem{ghz3} D. Bouwmeester, J.-W. Pan, M. Daniell, H. Weinfurter and A.
 Zeilinger, {\it Observation of Three-Photon Greenberger-Horne-Zeilinger
 Entanglement}, Phys. Rev. Lett. {\bf 82}(1999) 1345-1349.

 \bibitem{nielson}
  M. Nielsen and I. Chuang, {\it Quantum Computation and Quantum Information}
 (Cambridge University Press, 1999).

 \bibitem{aharonov} D. Aharonov and M. Ben-Or, {\it Fault tolerant computation with
  constant error}, in {\em Proceedings of the Twenty-Ninth Annual ACM Symposium on the
  Theory of Computing}, pp. 176-188, 1997.

\bibitem{kitaev} A. Yu. Kitaev, {\it Fault-tolerant Quantum Computation By
 Anyons}, Ann. Phys. {\bf 303} (2003) 2. Arxiv: quant-ph/9712048.

 \bibitem{fklw} M. Freedman, A. Kitaev, M. Larsen and Z. Wang, {\it Topological
 Quantum Computation,} Bull. Amer. Math. Soc. (N.S.) \textbf{40}  no. 1 (2003)
 31--38.

\bibitem {kauffman} L.H. Kauffman, {\it Knots and Physics} (World Scientific
 Publishers, 2002).

 \bibitem{kl} L.H. Kauffman and S.J. Lomonaco Jr, {\it
 Q-Deformed Spin Networks, Knot Polynomials and Anyonic Topological Computation},
 J. Knot Theory Ramifications, {\bf 16} (2007) 267-332.
 Arxiv:  quant-ph/0606114.

 \bibitem{jones2} V. F. R. Jones, {\it Braid groups,
Hecke algebras and Type ${\rm II}\sb 1$ Factors,} Geometric methods
in operator algebras (Kyoto, 1983), 242--273, Pitman Res. Notes
Math. Ser., \textbf{123}, Longman Sci. Tech., Harlow, 1986.

\bibitem{jones3}  V. F. R. Jones, {\it Hecke Algebra Representations
of Braid Groups and Link Polynomials,} Ann. Math. \textbf{126}
(1987), 335--388.

\bibitem{DSetal} S. Das Sarma, M. Freedman, C. Nayak, S. Simon and A. Stern,
{\it Non-Abelian Anyons and Topological Quantum Computation},
preprint, ArXiv: cond-mat.str-el/0707.1889.

\bibitem{g} R. Griess, {\it Automorphisms of Extra Special Groups
and Nonvanishing Degree Two Cohomology}, Pacific J. Math.
\textbf{48} no. 2 (1973) 403--422.

\bibitem{cxg} J.L. Chen, Kang Xue and Mo-Lin Ge, {\it Berry Phase
and Quantum Criticality in Yang-Baxter Systems}, Ann. of Phys.
{\bf 323} (2008) 2614.

\bibitem{aravind}
 P.K. Aravind, {\it Borromean Entanglement of the GHZ state}, in
 {\em Potentiality, Entanglement and Passion-at-a-Distance}, Cohen,
 Robert S., Michael Horne, and John Stachel (eds.), Kluwer
 Academic Publishers, Boston 1997.

 \bibitem{abdm} B. Abdesselam, A. Chakrabarti, V.K. Dobrev and
 S.G. Mihov, {\it Higher Dimensional Unitary Braid Matrices:
 Construction, Associated Structures and Entanglements}, J.
 Math. Phys. {\bf 48} (2007) 053508; ArXiv:math/0702188.

\bibitem{bb} J.L. Brylinski and R. Brylinski, {\it Universal
Quantum Gates}, in {\em Mathematics of Quantum Computation},
Chapman \& Hall/CRC Press, Boca Raton, Florida, 2002 (edited by R.
Brylinski and G. Chen).

\bibitem{fujii} K. Fujii, H. Oike and T. Suzuki, {\it More on the
Isomorphism $SU(2)\otimes SU(2)\cong SO(4)$}, Int. J. Geom. Methods
Mod. Phys. {\bf 4} no. 3 (2007) 471--485.


\bibitem{bonetal} N. Bonesteel, L Hormozi, G. Zikos, S. Simon, \emph{Braid
Topologies for Quantum Computation.}  Phys. Rev. Lett. \textbf{95}
(2005),  no. 14, 140503, 4 pp.

\bibitem{crss1} A. Calderbank, E. Rains, P. Shor and N. Sloane, {\it
Quantum Error Correction and Orthogonal Geometry}, Phys. Rev. Lett.
\textbf{78} (1997) 405.

\bibitem{crss2} A. Calderbank, E. Rains, P. Shor and N. Sloane, {\it
Quantum Error Correction via Codes over GF(4)}, IEEE {\it Transactions
on Information Theory}, \textbf{44(4)} (1998) 1369.

\bibitem{gottesman} D. Gottesman, {\it Stablizer Codes and Quantum Error Correction},
 Ph.D. Thesis, Caltech, 1997.

\bibitem{flw} M. Freedman, M. Larsen, Z. Wang,
 {\it The Two-Eigenvalue Problem and Density of Jones Representation
 of Braid Groups,} Comm. Math. Phys. \textbf{228} (2002) 177-199.
  ArXiv: math.GT/0103200.

\bibitem{GJ} D.M. Goldschmidt and V.F.R. Jones,
 {\it Metaplectic Link Invariants}, Geom. Dedicata \textbf{31} (1989) 165-191.

\bibitem{oster2000} A. Osterloch, L. Amico, U. Eckern, \emph{Bethe
Ansatz Solution of a New Class of Hubbard-type Models.} J. Phys.
{\bf A33} (2000) L87; F\emph{ermionic Long-range Correlations Realized
by Particles Obeying Deformed Statistics.} J. Phys. {\bf A33}
(2000) L487; \emph{Exact Solution of Generalized Schulz-Shastry type
Models.} Nucl. Phys. {\bf B 588} (2000) 531.

\bibitem{oster2009} A. Osterloh, J. Siewert, \emph{The Invariant-Comb Approach
and Its Relation to the Balancedness of Multipartite Entangled
States.} ArXiv: 0908.3818.

\end{thebibliography}
\end{document}